\documentclass[prc,preprint]{revtex4}

\usepackage{graphicx}

\begin{document}

\title{\bf Hot Spin
Polarized Strange Quark Stars in the Presence of Magnetic Field using a density dependent bag constant}

\author{{ G. H. Bordbar$^{1,2}$
\footnote{Corresponding author. E-mail:
bordbar@physics.susc.ac.ir}} and { Z. Alizade}}
\affiliation{$^1$Department of Physics,
Shiraz University, Shiraz 71454, Iran\\
and\\$^2$Center for Excellence in Astronomy and Astrophysics (CEAA-RIAAM)-Maragha,
P.O. Box 55134-441, Maragha 55177-36698, Iran}
\begin{abstract}
 The effect of magnetic field on the structure properties of hot spin polarized strange quark stars has been investigated.
For this purpose, we use the MIT bag model with a density dependent bag constant to calculate
the thermodynamic properties of spin polarized strange quark matter such as energy and equation of state.
We see that the energy and equation of state of strange quark matter changes significantly
in a strong magnetic field. Finally, using our equation of state, we compute the structure of
spin polarized strange quark star at different temperatures and magnetic fields.
\end{abstract}
\maketitle

 %%%%%%%%%%%%%%%%%%%%%%%%%%%%%%%%%%%%%%%%%%%%%%%%%%%%%%%%%%%%%%%%%%%%%%%%%%%%%%%%%%%%%
\section{Introduction}
A strange quark star is a hypothetical type of exotic star composed of strange quark matter. This is an ultra-dense phase of degenerate matter theorized to form inside particularly massive neutron stars.
It is theorized that when the degenerate neutron matter which makes up a neutron star is put under sufficient pressure due to the star's gravity, neutrons break down into their constituent up and down quarks. Some of these quarks may then become strange quarks and form strange matter, and hence a strange quark star, similar to a single gigantic hadron (but bound by gravity rather than the strong force).
Actually, until recently, astrophysicists were not sure there was a gray area between neutron stars and black holes, stellar remnants from a massive star's death had to be one or the other. Now, it is thought there is another bizarre creature out there, more massive than a neutron star, yet too small to collapse in on itself to form a black hole. Although they have yet to be observed, strange quark stars should exist, and scientists are only just beginning to realize how strange these things are.
Neutron  stars, strange quark stars and black holes are all born via the same mechanism: a supernova collapse. But each of them are  progressively more massive, so they originate from supernovae produced by progressively more massive stars. The collapsing  supernova will turn into a neutron star only if its mass is about $1.4-3 M_{sun}$. In a neutron star, if density of the core is high enough ($10^{15}\frac{gr}{cm^{3}}$) the nucleons dissolve to their components, quarks, and a hybride star (neutron star with a core of strange quark matter (SQM)) is formed. If after the explosion of the supernova density high enough ($10^{15}\frac{gr}{cm^{3}}$), the pure strange quark star (SQS) may be formed directly.
The composition of SQS was first proposed by Itoh \cite{rk1} with formulation of Quantum Charmo Dynamics (QCD).

One of the most important characteristics of a compact star is its magnetic field which is about $10^{15}-10^{19}\ G$  for pulsars, magnetars, neutron stars and SQS \cite{rk1017,rk1018}. This strong magnetic field has an important influence on compact stars. Therefore, investigating the effect of an strong magnetic field on strange quark matter (SQM) properties is important in astrophysics.
In recent years much interesting work has been done on the properties of dense astrophysical matter in the presence of a strong magnetic field \cite{rk2,rk3}. The effect of the strong magnetic field on SQM has been investigated using the MIT bag model as well as the D3QM model of confinement \cite{rk4,rk5}. We have studied the effects of strong magnetic fields on the neutron star structure employing the
lowest order constrained variational technique \cite{rk5-1}. Recently, we have also calculated the structure of polarized SQS at zero temperature \cite{rk6}, the structure of unpolarized SQS at finite temperature \cite{rk7}, structure of the neutron star with the quark core at zero temperature \cite{rk8} and finite temperature \cite{rk9, rk9p}, structure of spin polarized SQS in the presence of magnetic field at zero temperature using density dependent bag constant \cite{rk1015} and at finite temperature using a fixed bag constant \cite{rk1016}.
 The aim of the present work is calculating some properties of polarized SQS at finite temperature in the presence of a strong magnetic field using the MIT bag model with a density dependent bag constant.
To this aim, in section \ref{II}, we calculate the energy and equation of state of SQM in the presence of magnetic field at finite temperatures by MIT bag model using a density dependent bag constant. Finally in section \ref{sec4}, we  solve the TOV equation, and calculate structure of SQS.

\section{Calculation of energy and equation of state of strange quark matter}
\label{II}
We study the properties of strange quark matter and resulting equation of state. The equation of state plays an important role in obtaining the structure of a star. From a basic point of view, the equation of state  for SQM should be calculated by Quantum chromodynamics (QCD). Previous researchers have investigated the properties of the strange stars using diffrent equations of state with interesting results  \cite{rk21,rk22,rk23}. There are many different models for deriving the equation of state of strange quark matter (SQM) such as MIT bag model \cite{rk10,rk11}, NJL model \cite{rk12,rk13} and perturbation QCD model \cite{rk14,rk15}. Here, we use MIT bag model using a density dependent bag constant to calculate the equation of state of SQM in the presence of a strong magnetic field.

The MIT bag model confines three non-interacting quarks to a spherical cavity, with the boundary condition that the quark vector current vanishes on the boundary. The non-interacting treatment of the quarks is justified by appealing to the idea of asymptotic freedom, whereas the hard boundary condition is justified by quark confinement. This model developed in 1947 at "Massachusetts Institute of Technology". In this model quarks are forced by a fixed external pressure to move only inside a given spatial region and occupy single particle orbital. The shape of the bag is spherical if all the quarks are in ground state. Inside the bag, quarks are allowed to move quasi-free. It is an appropriate boundary condition at the bag surface that guarantees that no quark can leave the bag. This implies that there are no quarks outside the bag \cite{rk1020}.

\subsection{Density dependent bag constant}
  In the MIT bag model, the energy per volume for the strange quark matter is equal to the kinetic energy of the free quarks plus a bag constant $({ \cal B}_{bag})$ \cite{rk10}, which is the difference between energy densities of the noninteracting quarks and interacting quarks. There are two cases for the bag constant, a fixed value, and a density dependent value.
In the initial MIT bag model, two different values such as $55$ and  $90 \ MeV/fm^{3}$ were considered for the bag constant. Since the density of strange quark matter increases from the surface to the core of a strange quark star, it is more realistic that we use a density dependent bag constant \cite{rk1000,rk1001,rk1002,rk1003}. By considering the experimental date received at CERN, the quark-hadron transition occurs at a density about seven times the normal nuclear matter energy density $(156 \ MeV/fm^{3})$ \cite{rk15,rk1004}. By supposing that transition of quark-gluon plasma is only defined by the value of the energy density, the density dependence of ${ \cal B}_{bag}$ has been considered to have a Gaussian form,
\begin{equation}\label{001}
{{ \cal B}_{bag}}(n)={{\cal B}}_{\infty}+({\cal B}_{0}-{{\cal B}}_{\infty})e^{-\gamma(\frac{n}{n_{0}})^{2}},
\end{equation}
where ${\cal B}_{0}$ parameter is equal to ${\cal B}(n=0)$, and it has fixed value ${\cal B}_{0}=400 \ MeV/fm^{3}$. $\gamma$ is a numerical parameter, and usually equal to $0.17$, the normal nuclear matter density \cite{rk1003}. ${\cal B}_{\infty}$ depends only on the free parameter ${\cal B}_{0}$.

For obtaining ${\cal B}_{\infty}$, we use the equation of state of the asymmetric nuclear matter, which should agree with empirical data.
For computing the equation of state of asymmetric nuclear matter, we apply the lowest order constrained variational (LOCV) many-body procedure
as follows \cite{,rk1005,rk1006,rk1007,rk1008,rk1009,rk1010,rk1011,rk1012,rk1013}.

The asymmetric nuclear matter is defined as a system consisting of $Z$ protons $(pt)$ and $N$ neutrons $(nt)$ with the total number density $n= n_{pt}+n_{nt}$ and proton fraction $x_{pt}=\frac{n_{pt}}{n}$, where $n_{pt}$ and $n_{nt}$ are the number densities of protons and neutrons, respectively. For this system, we consider a trial wave function as follows:
 \begin{equation}\label{002}
\psi=F\phi,
\end{equation}
where $\phi$ is the Slater determination of the single-particle wave function and F is the A-body correlation operator $(A \ = \ Z \ + \ N)$, which is taken to be \begin{equation}\label{003}
F=\textrm{S} \prod f(ij),
\end{equation}
and \textrm{S} is a symmetrizing operator. For the asymmetric nuclear matter, the energy per nucleon up to the two-body term in the cluster expansion is
 \begin{equation}\label{004}
E([f])=\frac{1}{A}\frac{<\psi|H|\psi>}{<\psi|\psi>}=E_{1}+E_{2}.
\end{equation}
The one-body energy, $E_{1}$, is
 \begin{equation}\label{005}
E_{1}=\sum\sum\frac{\hbar^{2}k_{i}^{2}}{2m},
\end{equation}
where labels $1$ and $2$ are used for the proton and neutron respectively, and $k_{i}$ is the momentum of particle $i$. The two-body energy, $E_{2}$, is
\begin{equation}\label{006}
E_{2}=\frac{1}{2A}\sum<ij|v(12)|ij-ji>,
\end{equation}
where
\begin{equation}\label{007}
v(12)=-\frac{\hbar^{2}}{2m}[f(12),[\nabla_{12}^{2},f(12)]+f(12)V(12)f(12).
\end{equation}
 $f(12)$ and $V(12)$ are the two-body correlation and nucleon-nucleon potential, respectively. In our calculations, we use $UV_{14}+TNI$ nucleon-nucleon potential \cite{rk1014}. The procedure of these calculations has been studied in \cite{rk1006}.
According to this discussion, we minimize the two-body energy with relation to the variations in the correlation function subject to the normalization constraint. From minimization of the two-body energy, we get a set of differential equations. We can compute the correlation function by numerically solving these differential equations. Finally, we get the two-body energy, and then the energy of asymmetric nuclear matter.

The empirical consequence at CERN acknowledge a proton fraction $x_{pt}=0.4$ (data are from probation accelerated nuclei) \cite{rk1003,rk1004}.Therefore to calculate ${\cal B}_{\infty}$, we use our results of the above formalism for the asymmetric nuclear matter characterized by a proton fraction $x_{pt}=0.4$.
According to the following method, the assumptions of the hadron-quark transition takes place at energy density equal to $1100 \ MeV/fm^{3}$ \cite{rk1003,rk1004}.
We find that the baryonic density of the nuclear matter is $n_{0}=0.98 \ fm^{-3}$ (transition density). At densities lower than this value, the energy density of the quark matter is higher than that of the nuclear matter. By increasing the baryonic density, these two energy densities become equal at the transition density, and above this value the nuclear matter energy density remains always higher.
Also, we determine ${\cal B}_{\infty}=8.99 \ MeV/fm^{3} $ by putting the energy density of the quark matter and that of the nuclear matter equal to each other.

\subsection{Energy of spin polarized strange quark matter at finite temperature in the presence of magnetic field}
In this section, we derive the EOS of SQM in the presence of magnetic field.
First, we calculate the energy of SQM. For this, we should find the quark densities in term of baryonic number density ($n_B$).
By imposing charge neutrality and chemical equilibrium (we suppose that neutrinons leave the system freely), we get the following relations \cite{rk15},
\begin{equation}\label{1}
\mu_{d} = \mu_{u}+\mu_{e},
\end{equation}
\begin{equation}\label{2}
\mu_{s} = \mu_{u}+\mu_{e},
\end{equation}
 \begin{equation}\label{3}
\mu_{s} = \mu_{d},
\end{equation}
\begin{equation}\label{4}
2/3 n_{u}-1/3 n_{s}-1/3 n_{d}-n_{e} = 0,
\end{equation}
where $\mu_{i}$ is the chemical potential and $n_{i}$ is the number density of quark $i$.
We can ignore the electrons ($n_{e}=0$) \cite{rk16,rk17,rk18}, and consider the strange quark matter (SQM) including u, d and s quarks.
Therefore, we have
\begin{equation}\label{5}
n_{u}=1/2(n_{s}+n_{d}).
\end{equation}

In the presence of the magnetic field, we have the spin polarized SQM including spin-up and spin-down u, d and s quarks.
Now, we introduce the polarization parameter as follows,
\begin{equation}\label{6}
\zeta_{i}=\frac{n_{i}^{+}-n_{i}^{-}}{n_{i}}.
\end{equation}
In the above equation, $n_{i}^{+}$ is the number density of spin-up quark $i$ and $n_{i}^{-}$ is the number density of spin-down quark $i$,
where $0\leq \zeta_{i}\leq 1$ and $n_{i}=n_{i}^{+}+n_{i}^{-}$.

The chemical potential $\mu_{i}$ for any value of the temperature ($T$) and number density ($n_{i}$) is obtained using the following constraint,
\begin{equation}\label{7}
n_{i}=\sum_{p=\pm}\frac{g}{2\pi^{2}}\int_{0}^{\infty} f(n_{i}^{(p)},k,T)k^{2}dk,
\end{equation}
where $g$ is degeneracy number of the system and
\begin{equation}\label{8}
f(n_{i}^{(p)},k,T)=\frac{1}{exp\left(\beta((m_{i}^{2}c^{4}+
\hbar^{2}k^2c^{2})^{1/2}-\mu_{i}(n_{i}^{(p)},T))\right)+1}
\end{equation}
is the Fermi-Dirac distribution function. In the above equation $\beta=1/k_{B}T$ and $m_{i}$ is the mass of quark $i$. It should be noted that in our calculations, we ignore the masses of u and d quarks, and we consider $m_{s}=150\ MeV$.

The energy of spin polarized SQM in the presence of the magnetic field within the MIT bag model is as follows,
 \begin{equation}\label{9}
\varepsilon_{tot}=\varepsilon_{u}+\varepsilon_{d}+\varepsilon_{s}+\varepsilon_{M}+{ \cal B}_{bag},
\end{equation}
where
\begin{equation}\label{10}
\varepsilon_{i}=\sum_{p=\pm}\frac{g}{2\pi^{2}}\int_{0}^{\infty}(m_{i}^{2}c^{4}+
\hbar^{2}k^{2}c^{2})^{1/2}f(n_{i}^{(p)},k,T)k^{2}dk.
\end{equation}
In our calculations, we suppose
that $\zeta=\zeta_{u}=\zeta_{d}=\zeta_{s}$. In Eq. (\ref{9}), ${ \cal B}_{bag}$ is the bag constant with a density-dependent value which has been introduced in  Eq. (\ref{001}), and $\varepsilon_{M}=\frac{E_{M}}{V}$  is the magnetic energy
density of SQM, where $E_{M}=-M.B$ is the magnetic energy.
If we consider the uniform magnetic field along $z$ direction, the contribution of magnetic energy of
the spin polarized SQM is given by
\begin{equation}\label{11}
E_{M}=-\sum_{i=u,d,s}M_{z}^{(i)}B,
\end{equation}
where $M_{z}^{(i)}$ is the magnetization of the system corresponding to particle $i$ which is given by
\begin{equation}\label{12}
M_{z}^{(i)}=N_{i}{\mu_{i}}\zeta_{i}.
\end{equation}
In the above equation, $N_{i}$ and ${\mu_{i}}$ are the number and magnetic moment of particle $i$, respectively (${\mu_{s}=-0.581\mu_{N}}$, $\mu_{u}=1.852\mu_{N}$ and $\mu_{d}=-0.972\mu_{N}$, where $\mu_{N}=5.05\times10^{-27} \ J/T $ is the nuclear magnetic moment \cite{rk0019}).
Finally, the magnetic energy density of spin polarized SQM can be obtained using the following relation,
\begin{equation}\label{13}
\varepsilon_{M}=-\sum_{i}n_{i}\mu_{i}\zeta_{i}B.
\end{equation}

We obtain the thermodynamic properties of the system using the Helmholtz free energy,
\begin{equation}\label{16}
F=\varepsilon_{tot}-TS_{tot},
\end{equation}
where $S_{tot}$ is the total entropy of SQM,
\begin{equation}\label{14}
S_{tot}=S_{u}+S_{d}+S_{s}.
\end{equation}
In Eq. (\ref{14}), $S_{i}$ is entropy of particle $i$,
\begin{eqnarray}\label{15}
S_{i}(n_{i},T) &=&-\sum_{p=\pm}\frac{3}{\pi^{2}}k_{B}\int_{0}^{\infty}[ f(n_{i}^{(p)},k,T)ln(f(n_{i}^{(p)},k,T)) \\ \nonumber
 &+& (1-f(n_{i}^{(p)},k,T))ln(1-f(n_{i}^{(p)},k,T))] k^{2}dk.
\end{eqnarray}

\subsection{Equation of state of spin polarized strange quark matter}
Equation of state of strange quark matter plays an important role in investigating the structure of strange quark star \cite{rk19,rk8,rk20}.
We can use the free energy to derive the equation of SQM in the presence of the
magnetic field with a density dependent bag constant, by the following relation,
\begin{equation}\label{17}
P=\sum_{i}(n_{i}\frac{\partial F_{i}}{\partial n_{i}}-F_{i}),
\end{equation}
where $P$ is the pressure of system and $F_i$ is the free energy of
particle $i$ .

\section{Results and discussion}
\label{sec4}

\subsection{Thermodynamic properties of spin polarized strange quark matter}

\begin{figure}
\includegraphics[width=0.8\textwidth]{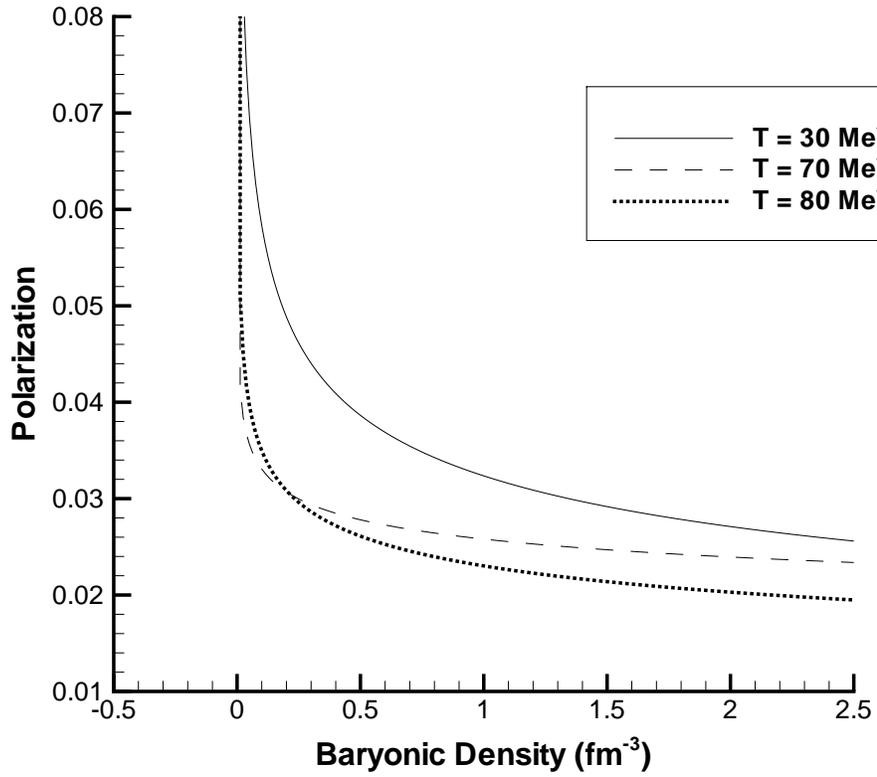}
\caption{The polarization parameter versus baryonic density  for $B=5\times10^{18} \ G$ at different temperatures $(T)$.}\label{fig1}
\end{figure}

\begin{figure}
\includegraphics[width=0.8\textwidth]{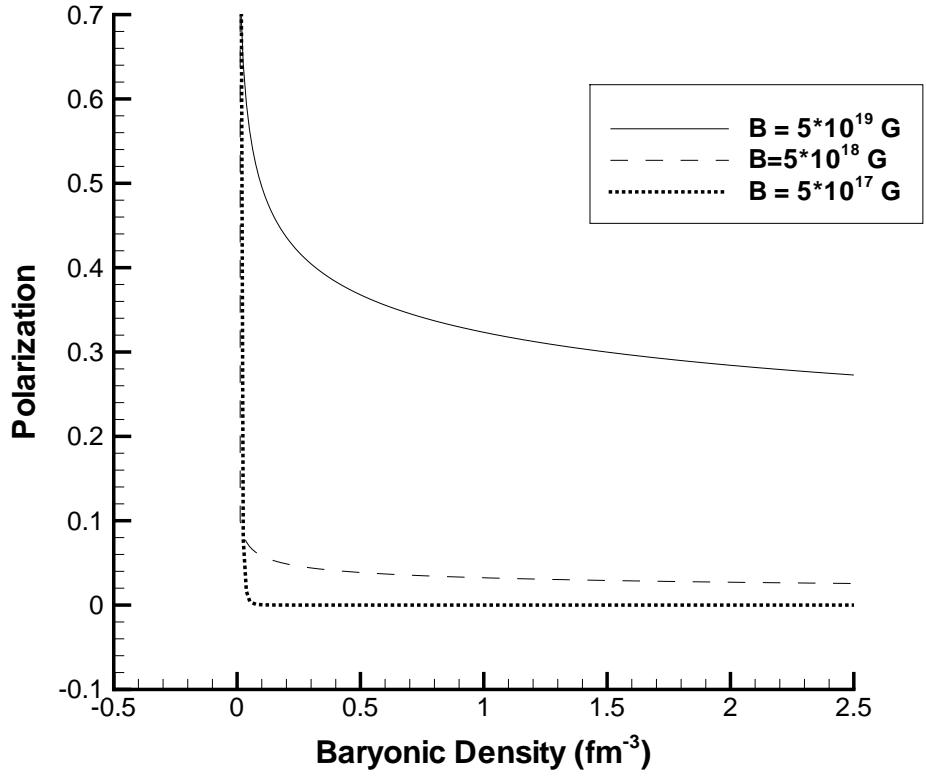}
\caption{The polarization parameter versus baryonic density at
$T=30 \ MeV$ for
different magnetic fields $(B)$.} \label{fig2}
\end{figure}

In Fig. \ref{fig1}, we have plotted the polarization parameter
versus the baryonic density in the presence of magnetic field ($B=5\times10^{18} \ G$)  at different temperatures.
From this figure, we can see that the polarization parameter decreases by increasing the baryonic density. However, at high  densities, the polarization parameter gets a constant value.
In Fig.\ref{fig1}, we have also shown the influence of increasing the temperature on the polarization of SQM. We see that at a fixed density, the polarization
parameter decreases by increasing the temperature. In fact, at high temperatures, the kinetic energy of quarks increases, and the contribution of
magnetic energy is therefore lower.
We have also shown the polarization parameter versus the baryonic density at a fixed temperature ($T=30\ MeV$) in
different magnetic fields in Fig. \ref{fig2}. This indicates that by increasing the baryonic density, the polarization parameter decreases. We see that at high densities, this parameter gets a constant value, and  it increases by increasing the magnetic field. Fig. \ref{fig2} shows that at high densities, for the magnetic fields lower than $B=5\times10^{17} \ G$, the polarization parameter  becomes nearly zero. In the other words, at high densities for low magnetic fields, the SQM becomes nearly unpolarized.

\begin{figure}
\includegraphics[width=0.8\textwidth]{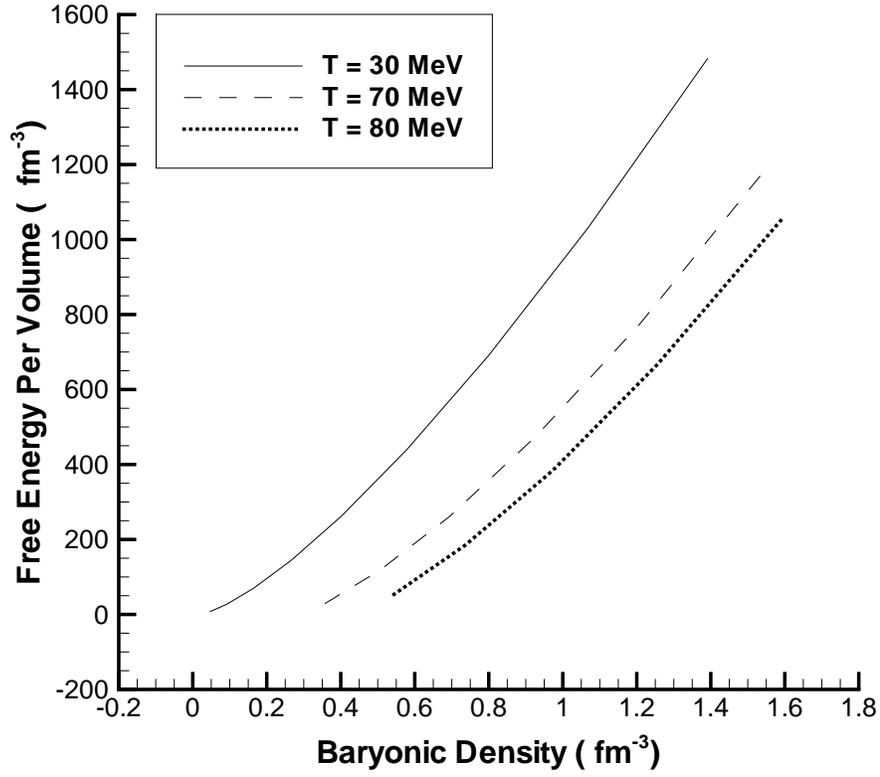}
\caption{The total free energy per volume of the spin polarized SQM as a function of the baryonic
density for $B=5\times10^{18}\ G$ at different temperatures $(T)$.} \label{fig3}
\end{figure}

\begin{figure}
\includegraphics[width=0.8\textwidth]{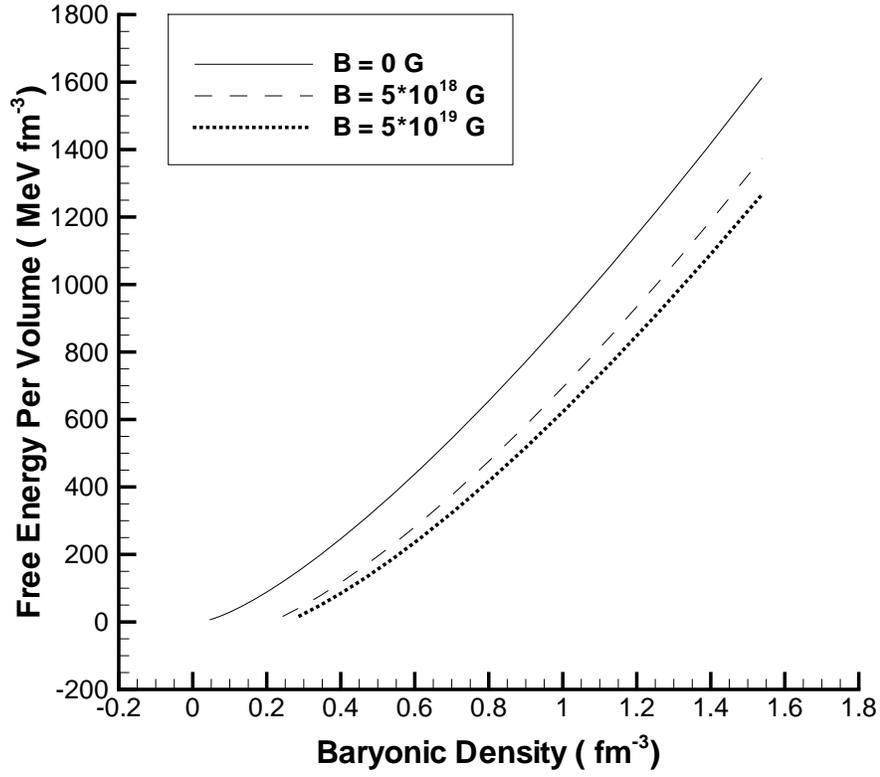}
\caption{The total free energy per volume of the spin polarized SQM as a function of the baryonic
density at $T=30 \ MeV$ for different magnetic fields $(B)$.} \label{fig4}
\end{figure}

We have presented the total free energy per volume of the spin polarized SQM  as a function of the baryonic density in  Fig. \ref{fig3}
for the magnetic field $B=5\times10^{18} \ G$ at different temperatures.
We can see that the free energy of spin polarized SQM increases by increasing the baryonic density,
and at high densities, the increasing of free energy is faster than at low densities.
At any density, the free energy decreases by increasing the temperature. This is due to the fact that the magnitude of second term of Eq. (\ref{16}) ($TS_{tot}$) increases as the temperature increases.
In Fig.  \ref{fig4}, we have seen that at a fixed temperature ($T=30 MeV$), the free energy of the spin polarized SQM decreases as
the magnetic field increases. In fact, the presence of magnetic field helps the orientation of quarks to a more regular and stable system with the lower energy.

\begin{figure}
\includegraphics[width=0.8\textwidth]{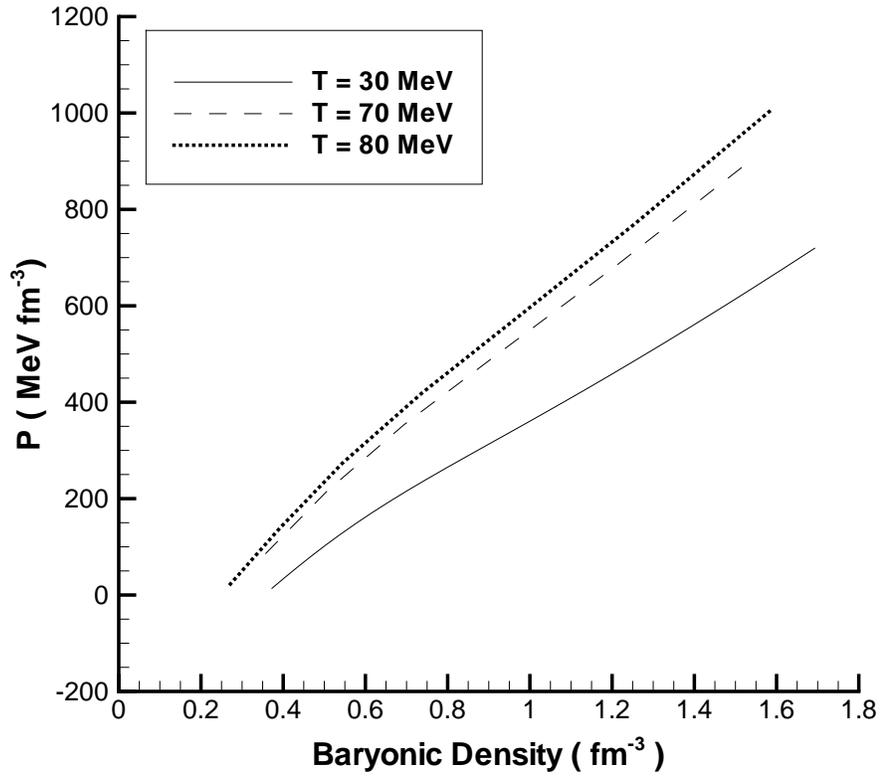}
\caption{The pressure of the spin polarized SQM versus the baryonic
density for $B=5\times10^{18}\ G$  at different temperatures $(T)$.} \label{fig5}
\end{figure}

\begin{figure}
\includegraphics[width=0.8\textwidth]{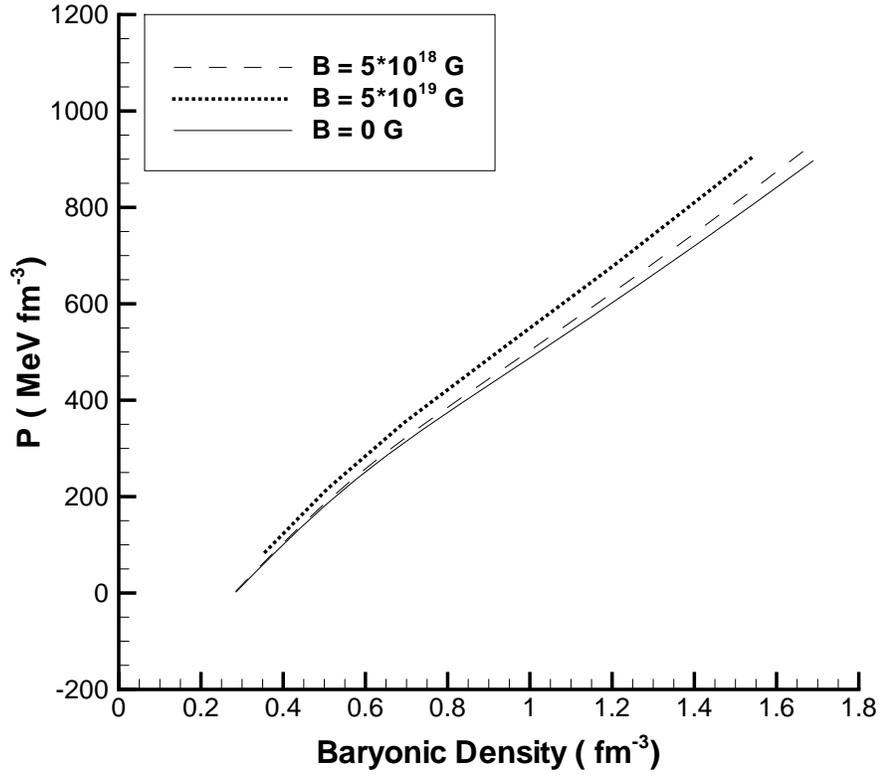}
\caption{The pressure of the spin polarized SQM the baryonic
density at $T=30 \ MeV$ for different magnetic fields $(B)$.} \label{fig6}
\end{figure}

\begin{figure}
\includegraphics[width=0.8\textwidth]{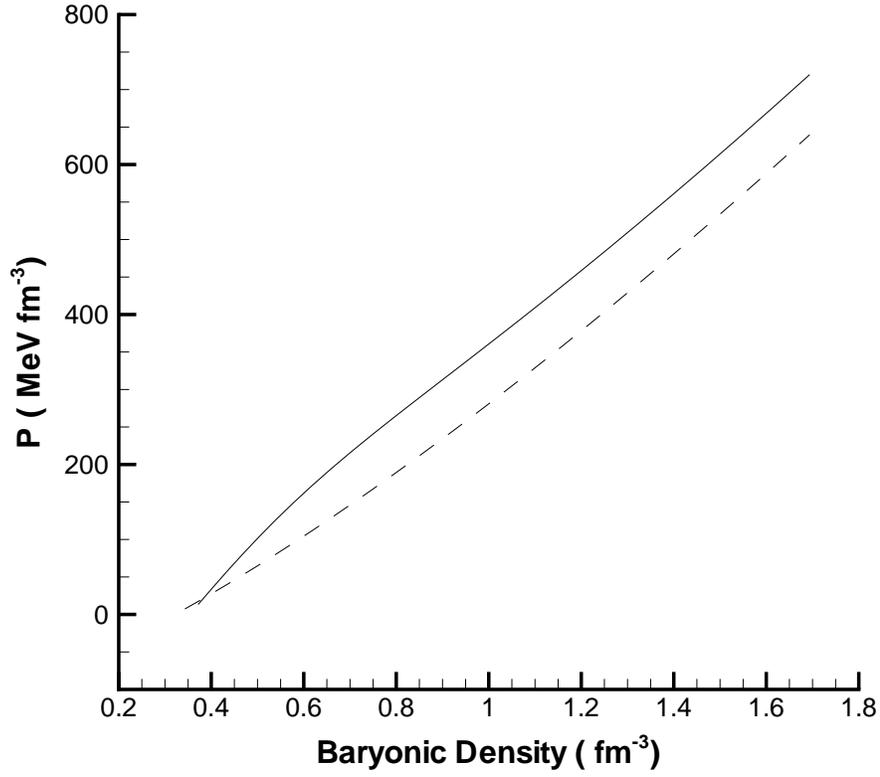}
\caption{The pressure of the spin polarized SQM the baryonic
density at $T=30 \ MeV$ and for $B=5\times10^{18}\ G$ calculated by a density dependent bag constant (solid  curve).
The results for ${ \cal B}_{bag}=90 \ MeVfm^{-3}$ (dashed  curve) have also been given for comparison.}  \label{fig7}
\end{figure}

In Fig. \ref{fig5}, we have shown the pressure of spin polarized SQM versus density in the presence of magnetic field ($B=5\times10^{18} \ G$) at different temperatures. From this figure, we have found that at each density, by increasing the temperature,
the pressure increases. In the other word, the equation of state of spin polarized SQM becomes stiffer by increasing the temperature. In Fig. \ref {fig6},
the equation of state of spin polarized SQM at fixed temperature ($T=30 MeV$) for different magnetic fields has been plotted. This figure indicates
that the presence of magnetic field leads to the stiffer equation of state for the spin polarized SQM.
As can be seen from Figs. \ref{fig3} and \ref{fig4}, by increasing both temperature and magnetic field,
increasing the free energy versus density takes place with the higher slope. This leads to higher pressure at higher temperatures and magnetic fields.
The equation of state of system for the density dependent bag constant at $T = 30\ MeV$ and $B=5\times10^{18} \ G$ has been plotted in Fig. \ref{fig7}. In this figure, we have also given the results for the case of fixed bag constant (${ \cal B}_{bag}=90\ \frac{MeV}{fm^{3}}$) \cite{rk1016} for comparison. Fig. \ref{fig7} indicates that with the density dependent bag constant, the equation of state of spin polarized SQM is stiffer than that with the fixed bag constant.

\subsection{Structure of spin polarized strange quark star}
\label{struc}

Mass and radius are the important macroscopic parameters for a compact star playing  crucial roles in investigation of its structure. Since strange quark stars are relativistic systems, for calculating the structure properties of these systems, we use general relativity.
We assume the strange quark star to be  spherically symmetric, the structure of this star is determined by numerically
integrating the Tolman-Oppenheimer-Volkoff equations \cite{rk24,rk25,rk26} using the equation of state of the system,
\begin{equation}\label{18}
\frac{dP}{dr}=-\frac{G\left[\varepsilon(r)+\frac{P(r)}{c^{2}}\right]\left[m(r)+\frac{4\pi
r^{3}P(r)}
{c^{2}}\right]}{r^{2}\left[1-\frac{2Gm(r)}{rc^{2}}\right]},
\end{equation}
\begin{equation}\label{19}
\frac{dm}{dr}=4\pi r^{2}\varepsilon(r),
\end{equation}
where $G=6.707\times10^{-45}\ MeV^{-2}$ is the gravitational constant, $r$ is the distance from the center of the star, $\varepsilon(r)$ is the energy density, $m(r)=m$ is the mass within the radius $r$, and $P=P(r)$ is the pressure. The boundary condition is $P(r=0)\equiv P_{c}=P(\varepsilon_{c})$, where $\varepsilon_{c}$ denotes the energy density at the star's center. For all pressure, we have $P<P_{c}$.

\begin{figure}
\includegraphics[width=0.8\textwidth]{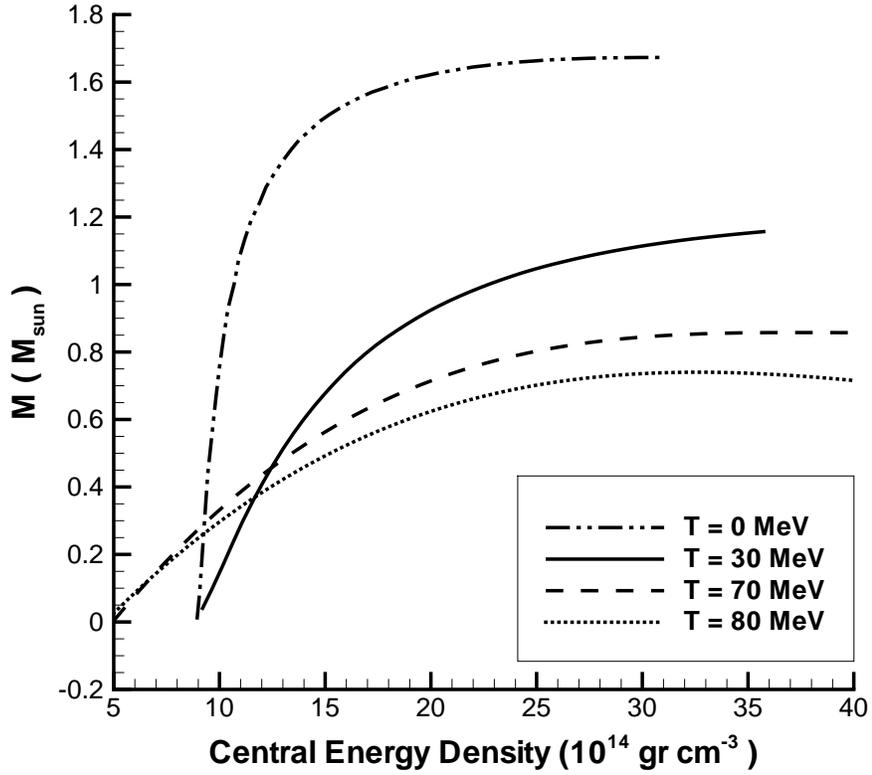}
\caption{The gravitational mass of spin polarized SQS versus the central energy density
in $B=5\times10^{18}\ G $ at different temperatures $(T)$. The results at $T=0 \ MeV$
(dashed dotted curve) have also been given for comparison.} \label{fig8}
\end{figure}

\begin{figure}
\includegraphics[width=0.8\textwidth]{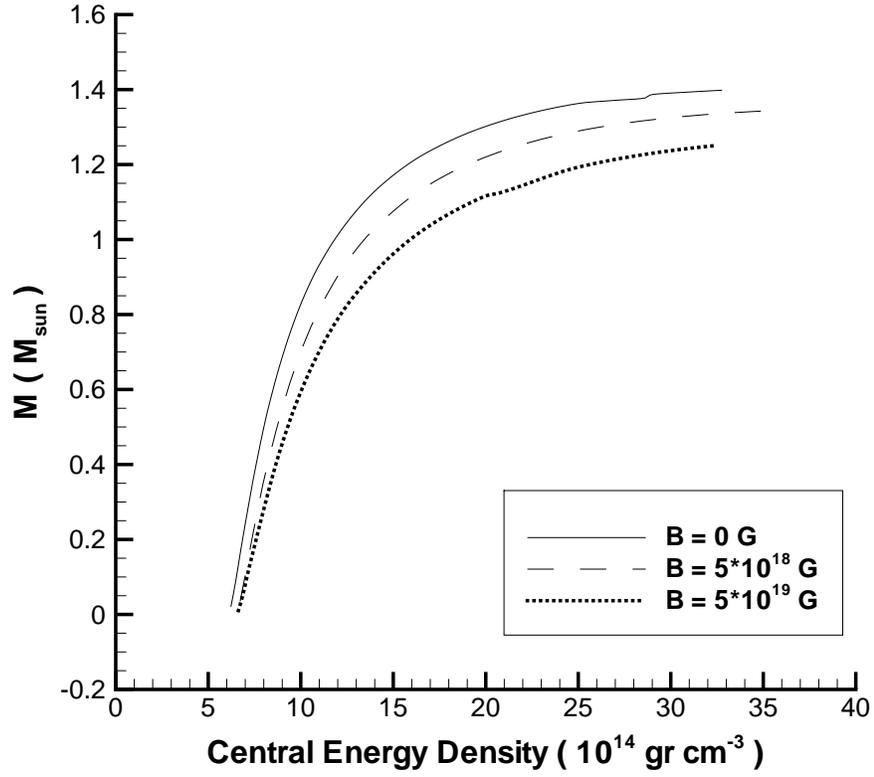}
\caption{The gravitational mass of spin polarized SQS versus the central energy density at $T=30 \ MeV$ for different magnetic fields $(B)$.}  \label{fig9}
\end{figure}

\begin{figure}
\includegraphics[width=0.8\textwidth]{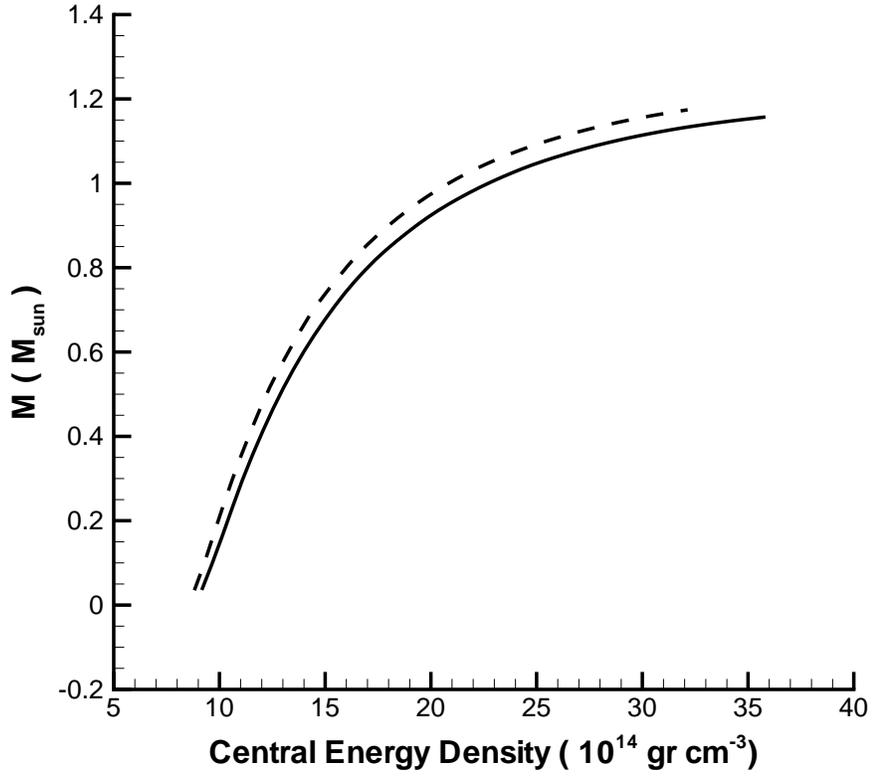}
\caption{The gravitational mass of spin polarized SQS versus the central energy density at $T= 30\ MeV$ for $B=5\times10^{18}\ G$ calculated by a density dependent bag constant (solid  curve).The results for ${ \cal B}_{bag}=90 \ MeVfm^{-3}$ (dashed  curve) have also been given for comparison.} \label{fig10}
\end{figure}

In Fig. \ref{fig8}, we have presented the gravitational mass of spin polarized SQS versus the central energy density at different temperatures for the magnetic field $B=5\times10^{18} \ G$. In this figure, we have also given the results at $T=0 \ MeV$ and $B=5\times10^{18} \ G$ for comparison \cite{rk1015}.
We can see that for all temperatures, the gravitational mass
increases rapidly by increasing the central energy density, and finally gets a limiting value (maximum gravitational mass).
This limiting value decreases by increasing the temperature.
The effect of  magnetic field on the gravitational mass of spin polarized SQS at a fixed temperature $T=30 \ MeV$ has been shown in Fig. \ref{fig9}.
We see that by increasing the magnetic field, the gravitational mass decreases.
%
%The above results indicate that at higher temperatures and magnetic fields, an SQS can be stable if it has a lower gravitational mass.
%
%
In Table \ref{T1}, we have given the maximum mass and the corresponding radius of spin polarized SQS at
different temperatures for $B=5\times10^{18} \ G$. It is shown that as the temperature increases,
the maximum mass and corresponding radius of spin polarized SQS decreases.
We have also presented the maximum mass and the corresponding radius of spin polarized
SQS for different magnetic fields at fixed temperature $T=30 \ MeV$ in Table \ref{T2}.
We see that the maximum mass and corresponding radius of the spin polarized SQS
decreases by increasing the magnetic field.
The above results indicate that at higher temperatures and magnetic fields, the spin polarized SQS with the lower gravitational mass can be stable.
From Figs. \ref{fig5} and \ref{fig6}, we see that by increasing the temperature and magnetic field, the equation of state of system becomes stiffer.
Here, we can conclude that the stiffer equation of state for spin polarized SQS leads to the lower values for its gravitational mass.
In Fig. \ref{fig10}, We have compared our results for two cases of density dependent and density independent bag constant (${ \cal B}_{bag}=90\ \frac{MeV}{fm^{3}}$) \cite{rk1016} at $T = 30\ MeV$ and $B=5\times10^{18} \ G$.
 We can see that in the case of density dependent bag constant, the gravitational mass of spin polarized SQS is lower than that in the case of fixed bag constant. This corresponds to the result of Fig. \ref{fig7} in which we have shown that the equation of state with the density dependent bag constant is stiffer than with the density independent bag constant.
In Table \ref{T3}, at $T=30 \ MeV$ for $B=5\times10^{18} \ G$, our results for the maximum mass and corresponding radius of spin polarized SQS has been compared with the results of density independent bag constant \cite{rk1016}. We can see that the maximum mass for the density dependent ${ \cal B}_{bag}$ is less than that for the fixed ${ \cal B}_{bag}$.

%
%================================================================================================
\section{Summary and conclusions}
In this article, we have studied the properties of a hot spin polarized strange quark matter (SQM) in the presence of the strong magnetic field by the MIT bag model using a density dependent bag constant.
We have shown that by increasing both magnetic field and temperature, the polarization parameter decreases. We have calculated the energy density and the equation of state of spin polarized SQM at different temperatures and magnetic fields. Our results show that by increasing both temperature and magnetic field, the  energy density decreases. It is seen that the equation of state of spin polarized SQM becomes stiffer by increasing both temperature and magnetic field.
We have used TOV equations to calculate the structure properties of spin polarized SQS. Our results show that the gravitational mass increases by increasing the central energy density and reaches a maximum value. This maximum value decreases by increasing both temperature and magnetic field.
From these results, we have concluded that at higher temperatures and magnetic fields, the SQS with lower gravitational mass can be stable.
We have compared our results of the density dependent bag constant with
results of a fixed bag constant.
It is  shown that the maximum mass with the density dependent bag constant is lower than that with a fixed bag constant.
%%%%%%%%%%%%%%%%%%%%%%%%%%%%%%%%%%%%%%%%%%%%%%%%%%%%%%%%%%%%%%%%%%%%%%%%%%%%%%%%%%%%%%%%%%%%%%%%%%
\section*{Acknowledgements}
{This work has been supported financially by the Center for Excellence in
Astronomy and Astrophysics (CEAA-RIAAM). We wish to thank the Shiraz University
Research Council.}

%%%%%%%%%%%%%%%%%%%%%%%%%%%%%%%%%%%%%%%%%%%%%%%%%%%%%%%%%%%%%%%%%%%%%%%%%%%%%%%%%%%%%%%%%%%%%%%%%%

%%%%%%%%%%%%%%%%%%%%%%%%%%%%%%%%%%%%%%%%%%%%%%%%%%%%%%%%%%%%%%%%%%%%%%%%%%%%%%%%%%%%%%%%%%%%%%%%%%%%%%

%%%%%%%%%%%%%%%%%%%%%%%%%%%%%%%%%%%%%%%%%%%%%%%%%%%%%%%%%%%%%%%%%%%%%%%%%%%%%%%%%%%%%%%%
\newpage
\begin{table}[h]
\begin{center}
  \caption[]{Maximum mass and the corresponding radius of spin polarized SQS for
  $B = 5 \times 10^{18}\ G$ at different temperatures.
 The results of  $T=0\ MeV$  have been
  also given for comparison \cite{rk1015}.}\label{T1}
  \begin{tabular}{clclclclcl}
  \hline\noalign{\smallskip}
 $T\ (MeV)$& & $M_{max}\ (M_{\odot})$& & $R\ (km)$  \\
 \hline\noalign{\smallskip}
 $0$ & &1.62 & & 8.36   \\
 $30$ & & 1.15 & & 7.1 \\
 $70$ & & 0.77 & & 6.89   \\
  \noalign{\smallskip}\hline
  \end{tabular}
\end{center}
\end{table}
%%=========================================================
\begin{table}[h]
\begin{center}
  \caption[]{Maximum mass and the corresponding radius of spin polarized SQS for different
  magnetic fields at $T=30\ MeV$.}\label{T2}
  \begin{tabular}{clclclcl}
  \hline\noalign{\smallskip}
 $B\ (G)$ & & $M_{max}\ (M_{\odot})$ & & $R\ (km)$  \\
 \hline\noalign{\smallskip}
 $0$ & & 1.39 & & 8.5 \\
 $5\times10^{18}$ & & 1.15 & & 7.1   \\
 $5\times10^{19}$ & & 0.99 & & 7.09 \\
 \noalign{\smallskip}\hline
  \end{tabular}
\end{center}
\end{table}
%%%%%%%%%%%%%%%%%%%%%%%%%%%%%%%%%%%%%%%%%%%%%
%%==================================================
\begin{table}[h]
\begin{center}
  \caption[]{Maximum mass and the corresponding radius of spin polarized SQS for
  $B = 5 \times 10^{18}\ G$ at $T=30\ MeV$.
 The results of  $T=30\ MeV$ by a fixed bag constant have been
  also given for comparison \cite{rk1016}.}\label{T3}
  \begin{tabular}{clclclclcl}
  \hline\noalign{\smallskip}
 ${ \cal B}_{bag}\ (MeV/fm^{3})$& & $M_{max}\ (M_{\odot})$ && $R\ (km)$  \\
 \hline\noalign{\smallskip}
  density dependent &   & 1.15 & &7.1   \\
 90  & & 1.17 && 7.37 \\
  \noalign{\smallskip}\hline
  \end{tabular}
\end{center}
\end{table}
%%=================================================================
%%%%%%%%%%%%%%%%%%%%%%%%%%%%%%%%%%%%%%%%%%%%%%%%%%%%%%%%%%%%%%%%%%%%%%%%%%%%%%%%%%%%%%%%%%%%%%%%%%%%%%
\newpage
%%%%%%%%%%%%%%%%%%%%%%%%%%%%%%%%%%%%%%%%%%%%%%%%%%%%%%%%%%%%%%%%%%%%%%%%%%%%%%%%%%%%%%%%%%%%%%%%%%%%%%%
%%%%%%%%%%%%%%%%%%%%%%%%%%%%%%%%%%%%%%%%%%%%%%%%%%%%%%%%%%%%%%%%%%%%%%%%%%%%%%%%%%%%%%%%%%%%%%%%%%%%%%%%%%
\newpage
%%%%%%%%%%%%%%%%%%%%%%%%%%%%%%%%%%%%%%%%%%%%%%%%%%%%%%%%%%%%%%%%%%%%%%%%%%%%%%%%%%%%%%%%%%%%%%
%%%%%%%%%%%%%%%%%%%%%%%%%%%%%%%%%%%%%%%%%%%%%%%%%%%%%%%%%%%%%%%%%%%%%%%%%%%%%%%%%%%%%%%%%%%%%%
\newpage
%%%%%%%%%%%%%%%%%%%%%%%%%%%%%%%%%%%%%%%%%%%%%%%%%%%%%%%%%%%%%%%%%%%%%%%%%%%%%%%%%%%%%%%%%%%%%
%%%%%%%%%%%%%%%%%%%%%%%%%%%%%%%%%%%%%%%%%%%%%%%%%%%%%%%%%%%%%%%%%%%%%%%%%%%%%%%%%%%%%%%%%%%%%%
\newpage
%%%%%%%%%%%%%%%%%%%%%%%%%%%%%%%%%%%%%%%%%%%%%%%%%%%%%%%%%%%%%%%%%%%%%%%%%%%%%%%%%%%%%%%%%%%%%
%%%%%%%%%%%%%%%%%%%%%%%%%%%%%%%%%%%%%%%%%%%%%%%%%%%%%%%%%%%%%%%%%%%%%%%%%%%%%%%%%%%%%%%%%%%%%%
\newpage
%%%%%%%%%%%%%%%%%%%%%%%%%%%%%%%%%%%%%%%%%%%%%%%%%%%%%%%%%%%%%%%%%%%%%%%%%%%%%%%%%%%%%%%%%%%%%%%%%%%
\newpage
%%%%%%%%%%%%%%%%%%%%%%%%%%%%%%%%%%%%%%%%%%%%%%%%%%%%%%%%%%%%%%%%%%%%%%%%%%%%%%%%%%%%%%%%%%%%%%
\newpage
%%%%%%%%%%%%%%%%%%%%%%%%%%%%%%%%%%%%%%%%%%%%%%%%%%%%%%%%%%%%%%%%%%%%%%%%%%%%%%%%%%%%%%%

%%%%%%%%%%%%%%%%%%%%%%%%%%%%%%%%%%%%%%%%%%%%%%%%%%%%%%%%%%%%%%%%%%%%%%%%%%%%%%%%%%%%%%%%%%%
%%%%%%%%%%%%%%%%%%%%%%%%%%%%%%%%%%%%%%%%%%%%%%%%%%%%%%%%%%%%%%%%%%%%%%%%%%%%%%%%%%%%%%%%%%%%%%%%%%%%%
\newpage
\end{document}